\documentclass[journal=jacsat,manuscript=article]{achemso}

\usepackage{chemformula} 
\usepackage[utf8]{inputenc}
\usepackage[T1]{fontenc} 

\usepackage[allcolors=blue,colorlinks]{hyperref}

\usepackage{color,soul}
\usepackage[version=4]{mhchem}

\author{Clàudia Climent}
\affiliation{Departamento de Física Teórica de la Materia Condensada and Condensed Matter Physics Center (IFIMAC), Universidad Autónoma de Madrid, E-28049 Madrid, Spain}
\email{claudia.climent@uam.es}

\author{David Casanova}
\affiliation{Donostia International Physics Centre (DIPC), 20018 Donostia, Euskadi, Spain}
\alsoaffiliation{IKERBASQUE, Basque Foundation for Science, 48009 Bilbao, Euskadi, Spain}

\author{Johannes Feist}
\affiliation{Departamento de Física Teórica de la Materia Condensada and Condensed Matter Physics Center (IFIMAC), Universidad Autónoma de Madrid, E-28049 Madrid, Spain}

\author{F. J. Garcia-Vidal}
\affiliation{Departamento de Física Teórica de la Materia Condensada and Condensed Matter Physics Center (IFIMAC), Universidad Autónoma de Madrid, E-28049 Madrid, Spain}
\alsoaffiliation{Institute of High Performance Computing, Agency for Science, Technology, and Research (A*STAR), Connexis, 138632 Singapore}

\title{Not dark yet: strong light-matter coupling can accelerate singlet fission dynamics}
\begin{document}

\begin{abstract}
Polaritons are unique hybrid light-matter states that offer an alternative way to manipulate chemical processes and change material properties. 
In this work we theoretically demonstrate that singlet fission dynamics can be accelerated under strong light-matter coupling. 
For superexchange-mediated singlet fission, state mixing speeds up the dynamics in cavities when the lower polariton 
is close in energy to the multiexcitonic triplet-pair state. We show that this effect is more pronounced in non-conventional singlet fission materials 
in which the energy gap between the bright singlet exciton and the multiexcitonic state is large ($> 0.1$ eV). In this case,
the dynamics is dominated by the polaritonic modes and not by the bare-molecule-like dark states, and additionally, the resonant enhancement due to strong coupling is very robust 
even for energetically broad molecular states.
The present results provide a new strategy to expand the range of suitable materials 
for efficient singlet fission by making use of strong light-matter coupling.
\end{abstract}
\newpage
\section{Introduction} 

The implications that strong light-matter coupling can have in chemistry have
recently raised a lot of interest. 
\cite{Ebbesen2016,Ribeiro2018,Feist2018,Hertzog2019a,herrera2020,Vidal2021}
This is because it offers an
unconventional way to manipulate chemical processes by modifying the energy
landscape as well as the dynamics. \cite{Hutchison2012b,Galego2015}
When an ensemble of molecules
interacts with a confined light mode, new eigenstates of the system emerge in
the strong coupling regime. \cite{Ribeiro2018,Feist2018}
This happens when the light-matter
interaction exceeds the intrinsic decay rates of both the molecular excitations
and the cavity photons. The new eigenstates of the system consist of two hybrid
light-matter states known as polaritons and a manifold of dark states,
superpositions of the molecular excitations that do not couple to the photon
mode. The main difference between both sets of states is that polaritons are
delocalized thanks to their cavity photon contribution, while dark states
usually behave similarly to uncoupled single-molecule excitons.

After a lot of progress in the field during the past decade, there is now a
solid understanding on the fundamentals of electronic strong coupling with molecules,
e.g., the modification of potential energy surfaces, 
\cite{Galego2015,Galego2016,Galego2017,Feist2018}
conical intersections 
\cite{Kowalewski2016Cavity,Vendrell2018b,Ulusoy2019,Csehi_2019,Cederbaum2021}
and electron and energy-transfer phenomena. 
\cite{herrera2016,doi:10.1063/1.5095940,PhysRevB.103.165412,Coles2014,Zhong2016,C8SC00171E,Saez-Blazquez2018Organic}
The finite lifetime of the cavity photons 
\cite{doi:10.1063/5.0033773,Antoniou2020,doi:10.1063/5.0011556,Felicetti2020}
and the presence of a dense dark state manifold 
\cite{PhysRevB.65.195312,PhysRevB.67.085311,Litinskaya2004,virgili2011,Herrera2017PRL,
Herrera2017PRA,Groenhof2019,https://doi.org/10.1002/adfm.202010737,Tichauer2021}
are key to understanding polaritonic chemistry
phenomena. As already noted, dark states may wash out polaritonic effects in
setups with collective, i.e., many-molecule, strong coupling, as there is a macroscopic number of them compared
to only two polaritons per cavity mode. \cite{cohenrisc,Vurgaftman2020,menon2020} 
For instance, recent work
has shown that there are limitations to strong coupling effects on
thermally activated delayed fluorescence (TADF), in which a triplet state
repopulates the singlet state responsible for the delayed
emission.\cite{karltadf,cohenrisc,doi:10.1063/1.5100192} This is because the initial state is
localized on a single molecule while a polariton is delocalized over the entire
molecular ensemble that couples to the cavity photon. As further discussed below, 
the rate for population relaxation from a
localized to a delocalized state is penalized by a factor $1/N$, where $N$ is
the number of entities participating in the delocalized state.
\cite{PhysRevB.67.085311,Litinskaya2004,delPino2015,joelsf}
In polariton-assisted TADF, direct population transfer from the triplet state to
the lower polariton then competes with the much faster process of relaxation between
localized single-molecule states, that is, from the triplet state to the singlet
dark state manifold. \cite{cohenrisc,doi:10.1063/1.5100192}
However, the reverse process is not penalized
when there is a manifold of $N$ localized final states, as the factor $1/N$ in
the rate to any single state is compensated by the number of states $N$.

Singlet fission is a downconversion photophysical reaction in which a
spin-singlet exciton splits into two independent spin-triplet states (equation
\ref{eq:sf_reaction}).\cite{Michl2010,michlann,Casanova2018} The recent interest
in this unique phenomenon has been driven by its potential capacity to overcome
the Shockley-Queisser limit \cite{SQlimit} for the efficiency of single junction
solar cells.\cite{nozik,congreve2013,baldosensitization} It is well accepted
that the singlet fission process involves the generation of a multiexcitonic
intermediate, the zero-spin triplet-pair state ($TT$),
\cite{Miyata2019,frutos2019} which eventually splits into two uncoupled triplet
states. Commonly, the formation of the $TT$ state (first step in equation
\ref{eq:sf_reaction}) is the rate-limiting process in singlet fission and is
thus the main subject of study in the field.
\begin{equation}
\mathbf{S_1} \ce{<=>[step 1]}  \mathbf{TT} \ce{<=>[step 2]} \mathbf{T_1} + \mathbf{T_1}
\label{eq:sf_reaction}
\end{equation}

In contrast to TADF under strong coupling, a polaritonic mode could potentially
be the initial state instead of the final one in a singlet fission process. Previous theoretical works investigating
singlet fission under strong coupling have mainly focused on the single-molecule
case, \cite{rubrene,Zhang2021,Gu2021} while studies of the collective situation
are still scarce.\cite{joelsf} There are no clear guidelines yet on if,
how, and when singlet fission can benefit from the presence of polaritons. This 
is the main motivation for the current work exploring cavity-modified singlet fission. We
focus on the situation of many-molecule strong coupling, which is
experimentally much easier to achieve than few-molecule strong coupling (only
possible in strongly subwavelength plasmonic nanocavities
\cite{Chikkaraddy2016}), and is simultaneously more promising for leading 
to practical light-harvesting devices in the strong-coupling
regime.\cite{Eizner2018,Nikolis2019,Saez-Blazquez2019,Esteso2021}. On the experimental side, there has been some initial work investigating singlet fission or
related processes such as triplet-triplet annihilation in optical cavities or
with plasmonic nanostructures. \cite{gomezrivas2019,musserchemsci,rubrene,menon2020,Ye2021}
Most of them have focused on the
long time (ns-$\mu$s) dynamics. \cite{musserchemsci,gomezrivas2019}  
In a recent work, transient
optical spectroscopy was employed to monitor the early dynamics of a TIPS-pentacene film
placed in a Fabry-Perot resonator. \cite{menon2020} The rate
constants extracted from fitting the experimental data did not show significant
differences between cavity and non-cavity situations. These results suggested
that the dark-state manifold dominated the dynamics and thus collective strong
coupling was unable to modify the singlet fission process.

In this work, we focus on singlet exciton fission dynamics under collective
strong coupling and explore how the dynamics is affected by the presence of dark
states and state-broadening due to both the 
natural linewidth of the vibronic peaks and energetic disorder. The central question we aim to answer
is whether the state mixing induced by strong light-matter coupling can enhance
the singlet fission rate in prototypical organic materials, and which materials are most suitable for such an application. 
Our results show that singlet fission 
indeed becomes faster when the lower polariton is spectrally tuned to be on resonance with the 
multiexcitonic state. Importantly, this mechanism is also operative for non-conventional 
singlet fission materials that present a large singlet-multiexcitonic state gap. We also find that 
the enhancement mechanism is not significantly affected by dark-state-induced dephasing in these compounds. 
Moreover, when energetic state broadening is considered, the enhancement in the rate 
is more robust against disorder for strongly exothermic materials than for conventional ones. 
This combination of properties opens up a whole range of
opportunities for materials that have not been explored for singlet fission to
date. Throughout the manuscript, we use italic characters to denote diabatic
electronic states, while eigenstates of the system are indicated in boldface and labeled according to their main diabatic contribution.

\color{black}
\section{Results and discussion}
\subsection{Singlet fission dynamics in an optical cavity}
To address the feasibility of singlet fission under strong light-matter
coupling, we first treat a single-molecule model and then extend it to the
many-molecule case in the following section. Our model describes several
molecular electronic excitations, a bath of intramolecular vibrations, and a
cavity photon. We fully consider the coupling between the electronic excitations and the 
photonic mode, and treat the
electronic-vibrational interaction perturbatively by relying on a master
equation approach based on Bloch-Redfield theory. There are two main reasons
behind this choice. First, it allows to naturally incorporate two essential
ingredients that have a leading role in light-matter strong coupling: the effect of many molecules as
well as the finite cavity lifetime (through an additional Lindblad term). Second,
in contrast to the commonly employed Fermi-Golden rule approaches that
rely on a perturbative treatment of the electronic couplings, 
\cite{Yost2014,doi:10.1063/1.4902135,doi:10.1063/1.4922644,Feng2014,Matsika2014}
Bloch-Redfield theory treats them exactly and can properly describe singlet
fission dynamics when the coupling to the vibrational bath is not too strong.
\cite{reichmanI, reichmanII, grozema2014}

The Bloch-Redfield master equation for the time evolution of the density matrix describing the electronic and photonic degrees of freedom
can be written as:
\begin{equation}
\frac{d}{dt}\rho_{ab}(t) = -i\omega_{ab} \rho_{ab}(t) + \sum_{c,d}R_{abcd}\rho_{cd}(t)
\label{eq:BR_eq}
\end{equation}
where $a,b,c$ and $d$ indices run over the eigenstates of the system
Hamiltonian, $\omega_{ab}$ are the eigenfrequency differences and $R_{abcd}$ is
the Bloch-Redfield tensor accounting for the system-bath interaction, i.e., the
coupling between electronic and vibrational modes. Relaxation rates are
expressed in terms of the spectral density representing a bilinear
electron-phonon interaction that we approximate by an Ohmic function with a
Lorentz-Drude cutoff, \cite{reichmanI, grozema2014} for which we have chosen
characteristic parameters of representative singlet fission materials. 
Although the specifics of the spectral density of the environment 
can be relevant for quantitative predictions for a given molecular
species, here we focus on providing insight and general guidelines on the
circumstances under which collective strong coupling modifies the singlet
fission dynamics, and thus use a general molecule-independent spectral density.

The system Hamiltonian for the molecule interacting with the cavity mode is
given by 
\begin{equation}
\hat H_S = \hat H_{el} + \hat H_{cav} + \hat H_{el-cav},
\label{eq:Hs}
\end{equation}
with the electronic Hamiltonian
\begin{equation}
\hat H_{el}=\sum_i E_i |i\rangle \langle i| + \sum_{i\neq j} V_{ij} |i\rangle \langle j|,
\label{eq:Hel} 
\end{equation}
where $E_i$ and $V_{ij}$ are the energies and interstate couplings,
respectively, for the (diabatic) electronic states involved in singlet fission.
We use a four-state model to represent a system with the ground state
($S_0$), the optically active singlet exciton ($S_1$), the triplet-pair state
($TT$) and a charge-transfer ($CT$) state, as suggested by Reichman and
coworkers. \cite{reichmanI} Specifically, we consider first the case of a slightly
exothermic formation of the $TT$ state ($E_{S_1}-E_{TT}=80$ meV,
$E_{TT}-E_{S_0}=1.7$ eV) and a higher-lying $CT$ state ($E_{CT}-E_{TT}=330$
meV). These energetics are representative of efficient singlet fission compounds
in which the $CT$ state mediates the population transfer from the $S_1$ to the
multiexcitonic $TT$ state via a superexchange mechanism. \cite{reichmanI} In
many singlet fission materials, the couplings between the $CT$ and the $S_1$ and
$TT$ states are (at least) one order of magnitude larger than the direct
$S_1/TT$ interaction, since the former contain one-electron terms while the
latter can be approximated as the difference between two bielectronic integrals.
\cite{Michl2010} Here we take $V_{S_1,CT}=V_{TT,CT}=30$ meV, whereas we
disregard the direct coupling ($V_{S_1,TT}=0$ meV). 

The cavity term in equation {\ref{eq:Hs}} takes the form 
$\hat H_{cav}= \hbar\omega_{c} \hat{a}^\dagger \hat{a}$, 
where $\omega_{c}$ is the frequency of the cavity mode, which we take in resonance with 
the optical exciton ($\hbar\omega_{c}=E_{S_1} - E_{S_0}$), and $\hat{a}^\dagger$ and $\hat{a}$ 
are the bosonic creation and annihilation operators, respectively.
Finally, the interaction between the cavity photon and the 
electronic states can be expressed by the Jaynes-Cummings Hamiltonian \cite{Jaynes1963} 
for the case of a single singlet fission site ($N=1$):
\begin{equation}
\hat H_{el-{cav}}= \sum_i \hbar g_i \big( \hat{a}^\dagger|S_0 \rangle\langle i| +  \hat{a}|i\rangle\langle S_0|\big),
\label{eq:Hel-cav}
\end{equation}
with $g_i$ being the coupling strength between the cavity photon and the $i$-th
electronic excited state, which is half the Rabi frequency ($\Omega_R$). Unless
otherwise indicated, we consider strong coupling to the bright state, with
$g_{S_1}=75$ meV. We precisely choose this value because it is roughly equal to
the energy gap between the $S_1$ and $TT$ states, thus placing the lower polariton ($\mathbf{LP}$)
close to the $\mathbf{TT}$ eigenstate. On the other hand, we consider both $TT$
and $CT$ to be optically non-active transitions, hence, their coupling to the
cavity vanishes. In this work we are interested in the linear-response regime,
such that simulations can be restricted up to the single-excitation subspace.

We account for the finite lifetime of the cavity photon due to radiative and
nonradiative decay by including a Lindblad term in our simulations. 
In this work we have considered a 50 ps
cavity lifetime, which can be achieved in dielectric cavities, 
even with deeply subwavelength mode volumes. \cite{doi:10.1126/sciadv.aat2355} 
When planning to manipulate molecular
photophysics with cavities, it is important to be aware of this additional
deactivation channel, which is absent in the non-cavity situation, and might
compete with the intrinsic molecular processes. In the situation we explore in
this work, the non-cavity singlet fission rate is faster than the cavity decay.
Bearing in mind that $\mathbf{TT}$ formation in pentacene, one of the most
efficient singlet fission materials, occurs within 80 fs, \cite{Wilson2011} 
one should aim for cavity lifetimes beyond the fs range
in order to modify singlet-fission dynamics via polariton formation.

\begin{figure}[H]
  \includegraphics[width=150mm]{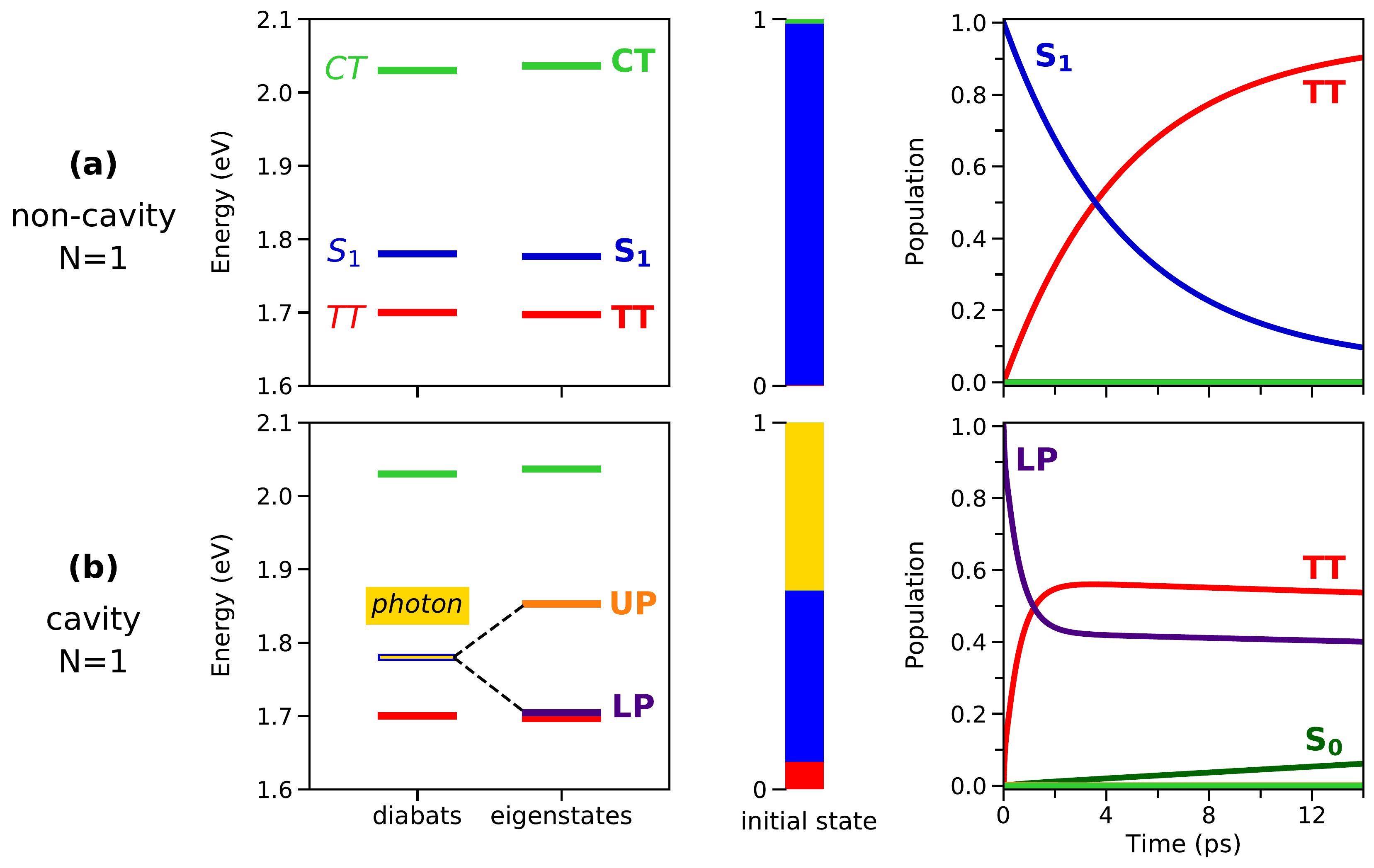}
  \caption{Energy spectrum (left), initial eigenstate diabatic composition (middle) 
  and population dynamics (right) for: 
  (a) Bare $N=1$ case, and strong coupling with 
  (b) $N=1$ ($g_{S_1}=75$ meV). 
  Note that the eigenstates are colored according to their main diabatic contribution. 
  Color code: cavity photon $|1\rangle$ (yellow), $S_0$ (dark green), $TT$ (red), $S_1$ (dark blue), $CT$ (lime green), 
  $\mathbf{LP}$ (indigo) and $\mathbf{UP}$ (orange).}
  \label{fig:sf_dyn}
\end{figure}

The singlet fission dynamics within and outside the optical cavity is represented
in \autoref{fig:sf_dyn}. For the non-cavity case (\autoref{fig:sf_dyn}a),
the initially populated $\mathbf{S_1}$ state relaxes to the $\mathbf{TT}$ state
with a mean time of 5 ps, as obtained from the Bloch-Redfield rate from the
adiabatic $\mathbf{S_1}$ to the $\mathbf{TT}$ eigenstate, 
\begin{equation}
  k_{\mathbf{S_1} \rightarrow \mathbf{TT}} =
  \big( \alpha_{TT}^2\beta_{TT}^2 + 
  \alpha_{S_{1}}^2\beta_{S_{1}}^2 + \alpha_{CT}^2\beta_{CT}^2 \big)
   S(\omega_{\mathbf{S_1},\mathbf{TT}}),
\label{eq:sf_rate}
\end{equation}
expressed in terms of the eigenstate expansion coefficients
$|\mathbf{S_1}\rangle=\sum_i\alpha_i|i\rangle$ and
$|\mathbf{TT}\rangle=\sum_i\beta_i|i\rangle $, and the power spectrum
$S(\omega)$, which characterizes the environment's ability to absorb or release
the energy required for the transition between the two eigenstates to happen at
a given finite temperature (in our case, room temperature). According to this
expression, the singlet fission rate will be non-zero under two conditions: i) both states must 
have a common diabatic contribution and ii) the environment of the molecular vibrations is able to meet the energetic
requirements for the transition to occur. As shown in \autoref{fig:sf_dyn}a,
the very small $TT$ and $CT$ contributions to the adiabatic $\mathbf{S_1}$ state are
sufficient to drive the population relaxation. Note that the time evolution of
the photophysical reaction proceeds with no significant population of the $CT$
state, which acts only as a mediator agent. 

Next, we focus on how strong light-matter coupling can affect the singlet fission dynamics. 
In this case, we explore the time evolution of the system 
starting from the $\mathbf{LP}$. 
Since the $\mathbf{LP}\rightarrow\mathbf{TT}$ population transfer rate is analogous 
to that for the non-cavity case in \autoref{eq:sf_rate} 
by replacing $\mathbf{S_1}$ with $\mathbf{LP}$, 
then a straightforward strategy to enhance the singlet fission rate might be 
to increase the mixing between states. For that, we couple the cavity photon 
with the $S_1$ state such that the $\mathbf{LP}$
is placed close to resonance with the $\mathbf{TT}$ state. 
In this scenario, the $\mathbf{LP}$ acquires some $TT$ character 
and the $\mathbf{TT}$ state acquires some polaritonic, i.e., $S_1$ and 
cavity photon, character, thus enhancing the $\mathbf{LP}\rightarrow\mathbf{TT}$ 
population transfer rate. 
This is indeed what happens, as shown in \autoref{fig:sf_dyn}b, 
where the singlet fission dynamics is accelerated at earlier times compared 
to the $\mathbf{S_1} \rightarrow \mathbf{TT}$ non-cavity situation,
with a considerably shorter mean time ($0.4$ ps). 

Although the rate populating the $\mathbf{TT}$ state is modified under strong
coupling, the CT-mediated singlet fission mechanism is maintained, as shown by
the nearly null population of the $CT$ state during the entire photophysical
reaction, and the fact that the multiexcitonic state is not populated without
the presence of the $CT$ state. Notice
also that because of the finite cavity lifetime (50 ps), the ground state ($S_0$)
becomes populated due to the non-zero cavity photon character of the
$\mathbf{LP}$ and $\mathbf{TT}$ states. It is important to highlight
that, since the $\mathbf{LP}$ is practically degenerate with the $\mathbf{TT}$
state, it is pure dephasing that promotes the population
transfer, i.e., it is the bath spectral density evaluated at zero frequency that determines the
Bloch-Redfield rate and, therefore, energy dependent details in this spectral density
are not too critical here.

\subsection{Collective strong light-matter coupling}
In the following we extend our model to the case of $N$ equivalent
(non-interacting) singlet fission sites, that is, multiple $\{S_1,TT, CT\}$
electronic systems, interacting with a cavity photon. For that, we expand the
electronic terms of the system Hamiltonian following the so-called Tavis-Cummings model
\cite{Tavis1968,Tavis1969}:
\begin{eqnarray}
\hat H_{el}=\sum_{k=1}^N\left[\sum_i E_i |i_k\rangle \langle i_k| + \sum_{i\neq j} V_{ij} |i_k\rangle \langle j_k| \right],
\label{eq:Hel_N} \\
\hat H_{el-cav}= \sum_i \hbar g_i^{(N)} \sum_{k=1}^N\big( \hat{a}^\dagger|S_{0_k} \rangle\langle i_k| +  \hat{a}|i_k\rangle\langle S_{0_k}|\big),
\label{eq:Hel-cav_N}
\end{eqnarray}
where $|i_k\rangle$ corresponds to an electronic excitation ($TT, S_1$ or $CT$)
at the $k$-th singlet fission site, and the light-matter coupling strength is
the same for all $k$ sites and is chosen so that the collective Rabi splitting
$\Omega_R$ stays constant ($g_i^{(N)}=\delta_{iS_1}\Omega_R/\sqrt{4N}$). In the
following we also restrict the system to one excitation at most.

\begin{figure}[H]
  \includegraphics[width=150mm]{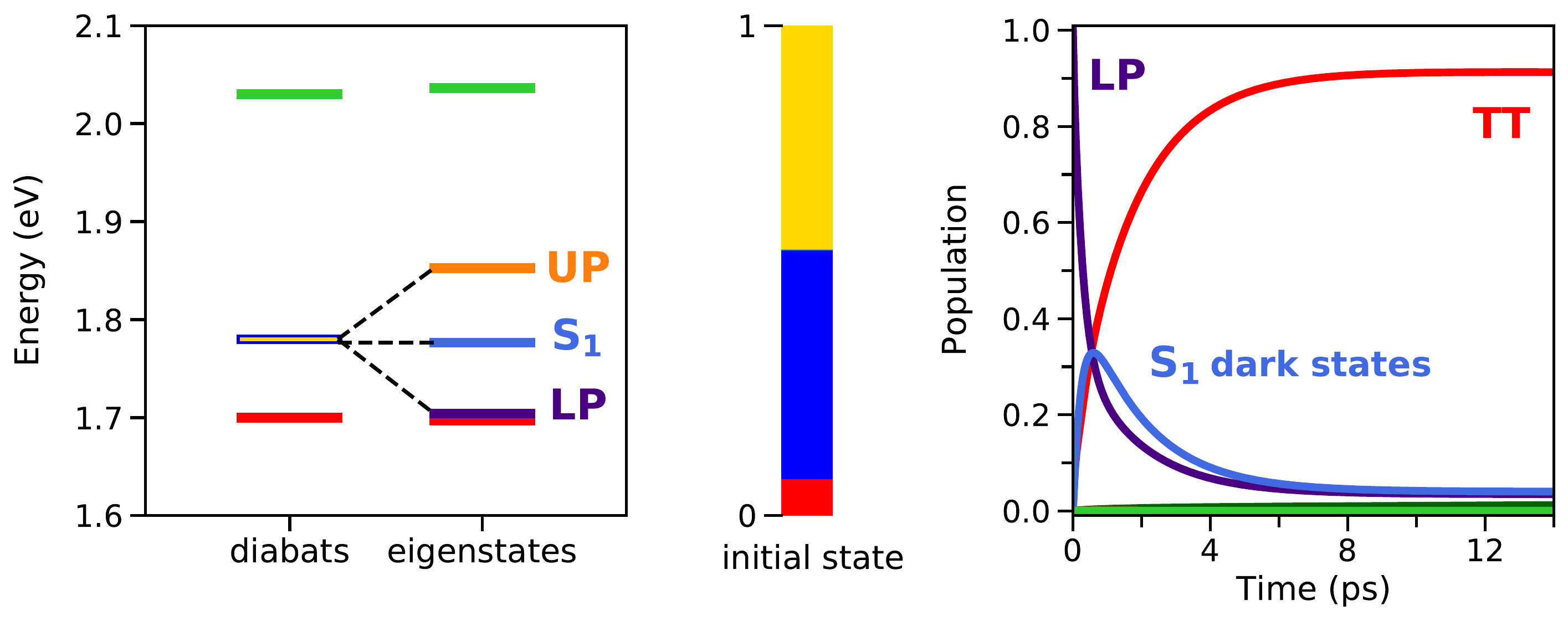}
  \caption{Energy spectrum (left), initial eigenstate diabatic composition (middle) 
  and population dynamics (right) for $N=20$ ($g_{S_1}=17$ meV). 
  Note that the eigenstates are colored according to their main diabatic contribution. 
  Color code: cavity photon $|1\rangle$ (yellow), $S_0$ (dark green), $TT$ (red), $S_1$ (dark blue), $CT$ (lime green), 
  $\mathbf{LP}$ (indigo), $\mathbf{UP}$ (orange), and $\mathbf{S_1}$ dark states (royal blue).}
  \label{fig:sf_dyn_N}
\end{figure}

The most important difference between the single singlet-fission site previously
discussed and the collective case is the presence of the dark singlet-exciton
manifold, i.e., linear combinations of the molecular $S_1$ states that do not
couple to the cavity photon. These states, together with the two polaritons and
the $\mathbf{TT}$ and $\mathbf{CT}$ manifolds, constitute the eigenstates of the
collective system. As shown in \autoref{fig:sf_dyn_N}, the fast
$\mathbf{LP} \rightarrow \mathbf{TT}$ relaxation is maintained, and because of 
detailed balance, the $\mathbf{TT}$ saturation population increases since for
larger $N$ the many $\mathbf{TT}$ states outnumber the lower polariton and act
as a population sink. Therefore, this polariton-enhanced mechanism
does not suffer from the issues arising in polariton-assisted TADF
\cite{karltadf,cohenrisc} previously discussed, in which a single-molecule excitation has to transfer to
a collective polariton, incurring a $1/N$ penalty in the transition rate.

It is straightforward to understand why the rapid initial relaxation from the
$\mathbf{LP}$ to the $\mathbf{TT}$ manifold still holds in the collective case.
The collective strong coupling version of \autoref{eq:sf_rate} is given by
\begin{equation}
  k_{\mathbf{LP} \rightarrow \mathbf{TT_q}} =
  \Big(\sum_k^N \alpha_{q,TT_k}^2\beta_{TT_k}^2 + 
  \alpha_{q,S_{1_k}}^2\beta_{S_{1_k}}^2 + \alpha_{q,CT_k}^2\beta_{CT_k}^2 \Big)
   S(\omega_{\mathbf{LP},\mathbf{TT_q}}),
\label{eq:sf_rate_N}
\end{equation}
where $q=1\dots N$ labels the specific state of the $\mathbf{TT}$ manifold.
These states are basically linear combinations of the non-cavity $\mathbf{TT}$
eigenstate on each site, $|\mathbf{TT}_q\rangle \approx \sum_{k}^{N}
\alpha_{q,TT_k} |TT_k\rangle$, therefore, the rate constant
can be approximated as $k_{\mathbf{LP}\rightarrow \mathbf{TT}_q} \approx \sum_k^N
\alpha_{q,TT_k}^2\beta_{TT_k}^2 S(\omega_{\mathbf{LP},\mathbf{TT}_q})$. Since
the $\mathbf{LP}{}$ amplitude of the $TT$ state on each $k$-site is equal to the
$N=1$ amplitude scaled by $1/N$, i.e., $\beta_{TT_k}=\beta_{TT}/\sqrt{N}$, and
since $\sum_k^N \alpha_{q,TT_k}^2 \approx 1$, then $k_{\mathbf{LP}\rightarrow
\mathbf{TT}_q} \approx \beta_{TT}^2 S(\omega_{\mathbf{LP},\mathbf{TT}_q})/N$.
Therefore, in the macroscopic limit, the rate from the $\mathbf{LP}$ to the set
of $N$ $\{\mathbf{TT}_q\}$ states is independent of $N$ and is dictated by the
amount of $TT$ character the $\mathbf{LP}$ acquires in the $N=1$ case
($\beta_{TT}$). According to this analysis, collective strong coupling will
impact the early singlet fission dynamics when the $\mathbf{LP}$ is close to
resonance with the $\mathbf{TT}$ manifold, such that $\beta_{TT} \neq 0$. Notice
that in the collective case, the ground state is barely populated in contrast to
the single-site results (\autoref{fig:sf_dyn}b). This is because only one
eigenstate of the $\mathbf{TT}$ manifold has a non-vanishing cavity
contribution.

\subsection{Promoting strongly exothermic singlet fission}
An important factor at play here that we have not addressed yet is the role of
the $(N-1)$ \{$\mathbf{S_{1}}_q$\} dark states. In the example we have discussed,
since this manifold is relatively close in energy to the $\mathbf{LP}$, it is
significantly populated during the dynamics, as shown in \autoref{fig:sf_dyn_N}. From
the $S_1$ dark states, singlet fission then proceeds with essentially the same
rate as in the non-cavity situation, then diminishing the rate enhancement due to strong coupling. 
However, the further these \{$\mathbf{S_{1}}_q$\} dark
states are spectrally separated from the $\mathbf{LP}$ and the $\mathbf{TT}$ manifold, the less
populated they will be after excitation to the $\mathbf{LP}$ and, as a consequence, 
they will be less detrimental to the cavity-based singlet-fission rate enhancement. 
Therefore, we hypothesize that the ideal candidates for singlet fission under collective strong
coupling are those compounds with a large enough $S_1-TT$ gap, such that when the
$\mathbf{LP}$ is excited, the \{$\mathbf{S_{1}}_q$\} dark states are energetically
too high to be populated. Standard singlet fission materials exhibit an $S_1$ state 
that ideally lies slightly above the
multiexcitonic $TT$ singlet state, i.e., weakly exothermic singlet fission, which 
severely limits the pool of potentially efficient singlet fission compounds.
Also, prototypical efficient singlet fission materials, such as pentacene and
rubrene, have poor photochemical stabilities, \cite{Maliakal2004,Mondal2009,Ly2018}
preventing their use in real-world devices. 
Therefore, according to our hypothesis, strong
coupling could enhance the singlet fission rate in materials that have usually
been ignored since excessive exoergicity is known to be detrimental in the
non-cavity situation. \cite{Zhang2016,C9SC05066C,Casanova2018}

\begin{figure}[H]
  \includegraphics[width=150mm]{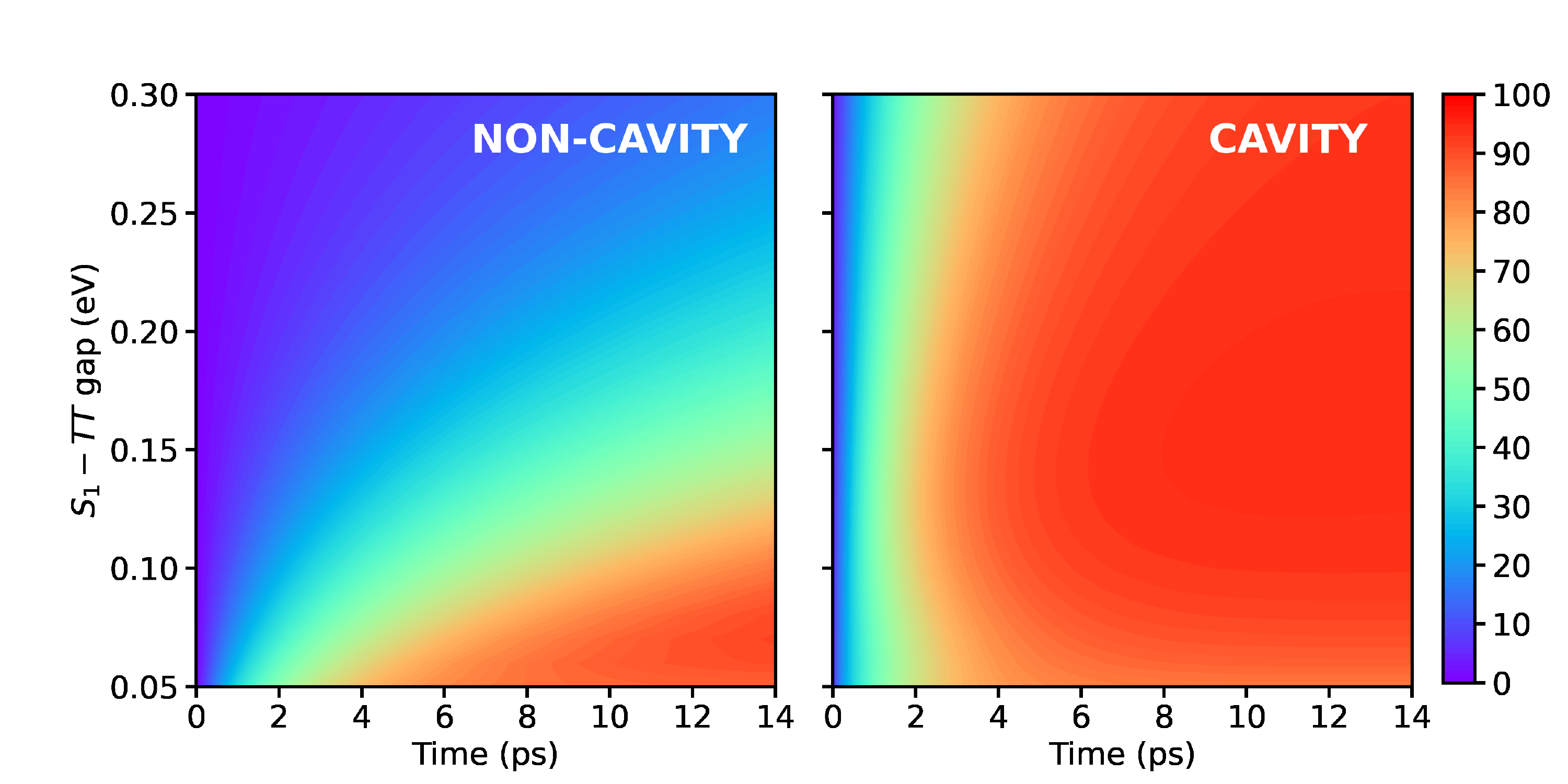}
  \caption{Adiabatic $\% \mathbf{TT}$ population for the non-cavity ($\mathbf{S_1}$ initial state) 
  and cavity ($\mathbf{LP}$ initial state and $N=20$) situations.
The single-molecule coupling constant $g_{S_1}=10-66$ meV as the $S_1-TT$ gap increases 
since a larger Rabi splitting is needed for the $\mathbf{LP}$ to be close to resonance 
with the $\mathbf{TT}$ eigenstates.}
\label{fig:tt_population}
\end{figure}

To test our hypothesis, in \autoref{fig:tt_population} we plot the
$\mathbf{TT}$ population dynamics dependence on the diabatic $S_1-TT$ gap while
keeping the $CT-S_1$ gap constant. For the cavity case, as the $S_1-TT$ gap
varies, the Rabi splitting is chosen such that the $\mathbf{LP}$ lies close to
resonance with the $\mathbf{TT}$ eigenstates ($5-10$ meV above), just like in
Figures \ref{fig:sf_dyn}b and \ref{fig:sf_dyn_N}. For the non-cavity case, the
singlet fission dynamics is greatly slowed down as the separation between the
two states increases as the amplitude contribution in \autoref{eq:sf_rate}
decreases due to reduced state-mixing. In contrast, within the cavity, the
$\mathbf{TT}$ state is populated much faster and independently of the $S_1-TT$
gap, as long as the $\mathbf{LP}$ is brought close to resonance with the
$\mathbf{TT}$ eigenstates. These results therefore indicate that strong
light-matter coupling can also accelerate the singlet-fission dynamics of
materials with large $S_1-TT$ gaps.

\begin{figure}[H]
  \includegraphics[width=100mm]{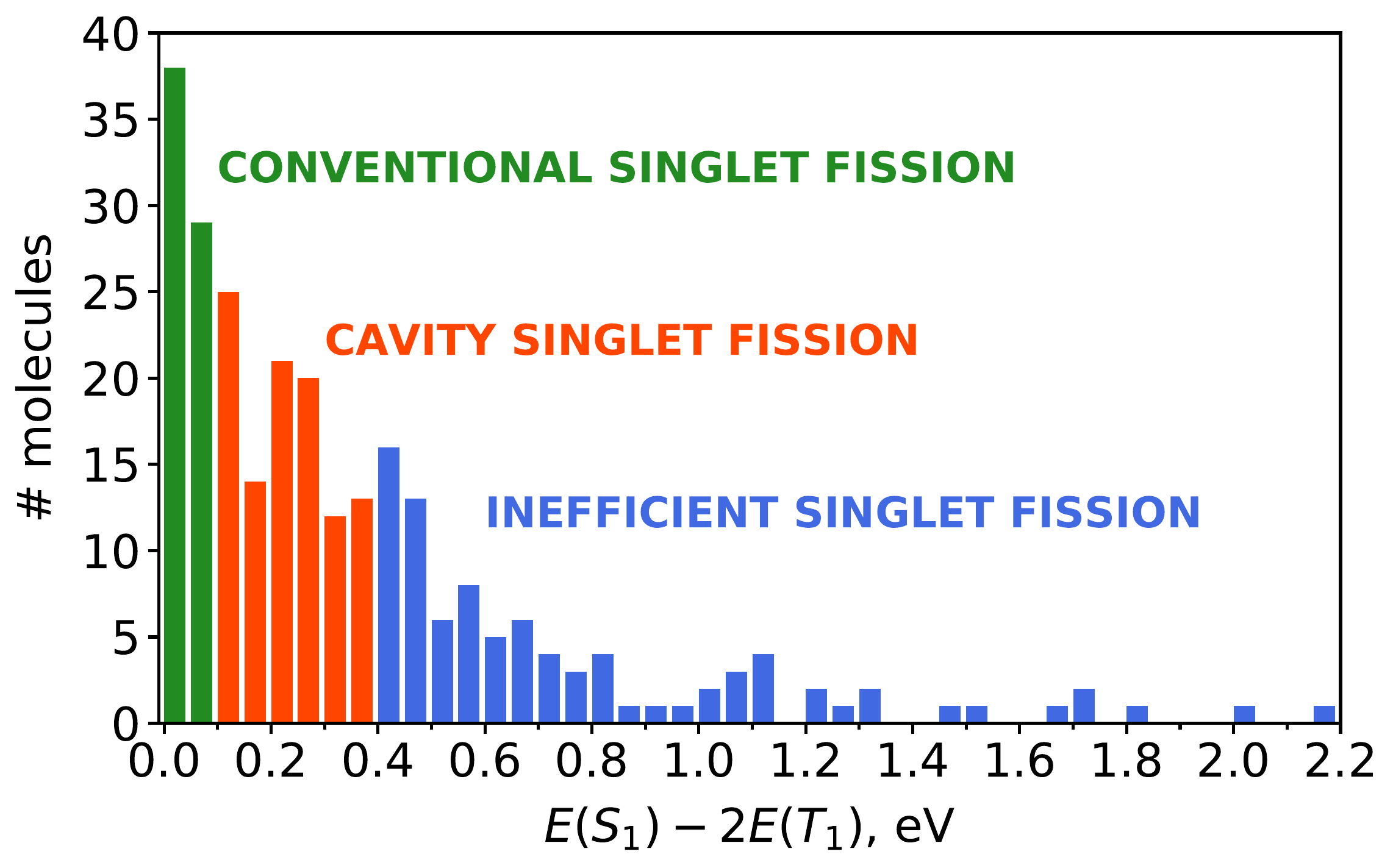}
  \caption{Histogram (in eV) of a list of 262 organic molecules. 
  Experimental structures from reference \citenum{Montalti2006}. 
Singlet and triplet energies obtained from reference \citenum{troisidatabase}.}
\label{fig:histogram}
\end{figure}

Therefore, within the strong coupling regime it becomes possible to efficiently
generate triplet-pair states not only for slightly exothermic energetics,
e.g., in pentacene. Relaxing the near degeneracy criterion
($E(S_1)-2E(T_1)\sim0$) should allow to obtain fast singlet fission processes in a wider
range of molecular systems with respect to those identified up to date. In
order to exemplify the potential impact in the search for singlet fission
chromophores, in the following, we consider a set of organic molecules with
suitable properties: (i) exothermic singlet fission ($E(S_1)-2E(T_1)>0$), 
(ii) sizable transition dipole moment between ground and lowest
excited singlet, so strong coupling to the cavity mode could be
achieved, and (iii) sufficiently large ($> 0.4$ eV) $S_0$-to-$T_1$ gap 
in order to ensure molecular stability for practical applications.  
Our molecular test set contains $262$ organic molecules obtained from the
Cambridge Structural Database (CSD) \cite{CSD} selected following the protocol
designed by Padula and collaborators \cite{troisidatabase}. 
In order for singlet fission
to compete with other relaxation pathways, such as internal conversion,
intersystem crossing or radiationless decay to the ground state, it should
take place on a ps (or even sub-ps) time scale. Therefore, here we consider that good
candidates for efficient singlet fission are those able to reach a 50\%
population of the $TT$ state within 5 ps after photoexcitation. Of course, this
is a somewhat arbitrary limit, but it reasonably serves our purpose to compare
singlet fission efficiency with and without strong coupling. For the non-cavity
scenario, it occurs for $E(S_1)-E(TT)\le0.1$ eV (\autoref{fig:tt_population}a),
which is in the order of the gap measured in crystalline pentacene,
\cite{pentacene1971,pentacenePRL} and slightly higher than the value employed in
the previous sections (0.08 eV). About 26\% of the considered chromophores
exhibit $E(S_1)-E(TT)\le0.1$ eV (green bars in Figure {\ref{fig:histogram}}),
and are expected to be the most promising singlet fission compounds. In
contrast, if we consider $\sim0.8$ eV as the upper limit for the Rabi splitting, \cite{Hutchison2012b,Cohen2013}
the number of potential molecules able to undergo singlet fission
efficiently when placed in an optical cavity increases to 66\% of the total set
of studied molecules (orange bars in \autoref{fig:histogram}). These results
demonstrate how strong coupling can substantially increase the pool of suitable
molecular candidates, which represents one of the main current challenges for
the practical implementation of singlet fission in optoelectronic devices.
\cite{Casanova2018} Also note that the values reported here can be seen as rather
conservative, since in the cavity model the $S_1-CT$ energy difference remains constant while
varying the $S_1-TT$ gap, debilitating the efficiency of the CT-mediated mechanism 
(for large $S_1-TT$ gaps, the CT states lie rather high with respect to $\mathbf{LP}$ and $\mathbf{TT}$). 
Moreover, we have not considered here the case of direct singlet fission, i.e., direct $S_1/TT$ coupling, 
which would also benefit from the energy level alignment between $\mathbf{LP}$ and $\mathbf{TT}$ states.

\subsection{Effect of state-broadening}
So far we have assumed equivalent singlet fission sites with well-defined
discrete $TT$, $S_1$ and $CT$ energies. Since our conclusions rely on the
$\mathbf{LP}$ being close to resonance with the $\mathbf{TT}$ manifold, it is
not clear whether they will still hold when taking into account 
the energetic broadening of the states. In the following, we account 
for the linewidth of the Franck-Condon vibronic transition that couples to the 
cavity mode and also inhomogeneous broadening due to energetic disorder 
because of different local environments.
To explore this effect, we select
the molecular energies by sampling a Gaussian distribution centered at $E_i$
with $i=TT,S_1,CT$. We consider a
maximum full-width half-maximum (FWHM) value of 0.1 eV, which is characteristic
of the $S_1$ vibronic absorption peaks of polycyclic aromatic hydrocarbons. \cite{doi:10.1063/1.3495764}
We explore two limiting cases, one in which the energies of the
three molecular states of each singlet fission site are jointly sampled, that
is, the same random number is used to sample the Gaussian distribution for all
three states of a given site, and another one for which the three states are
sampled independently by using different random numbers. Note that in the former
case, the energy gaps between states remain constant, while in the latter they
may vary. 

  \begin{figure}[H]
    \includegraphics[width=150mm]{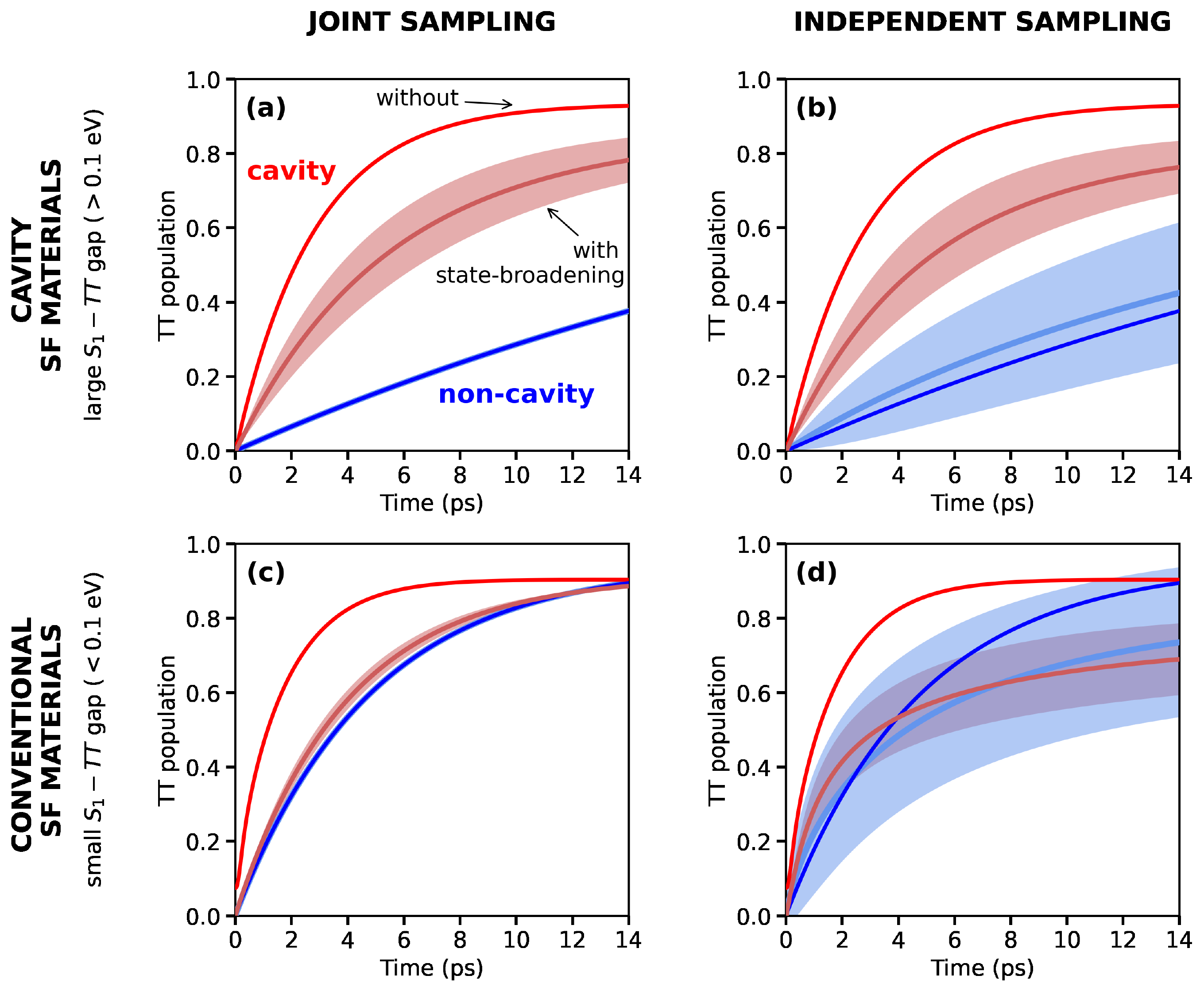}
    \caption{$TT$ (diabatic) population dynamics with (light) and without (dark) state broadening.
    Non-cavity results are plotted in blue and strong coupling ones in red. 
    The shaded area includes the mean $\pm$ standard deviation of 50 realizations. 
    Results are shown for the $N=20$ case.
    Coupling strength in (a),(b) $g_{S_1}=42$ meV, and in (c),(d) $g_{S_1}=17$ meV.
    The initial state for the cavity simulations is the $\mathbf{LP}$. 
    In the presence of state-broadening, it is taken to be 
    $|\mathbf{LP}\rangle = c_r|\mathbf{r}\rangle$, 
    where $c_r$ is the cavity-photon coefficient of the $r$-th eigenstate $|\mathbf{r}\rangle$, 
    and only those eigenstates below the cavity photon are considered.
    For the non-cavity simulations, the initial state is taken as the eigenstate with the 
    largest $S_1$ amplitude.}
    \label{fig:broadening}
  \end{figure} 

In \autoref{fig:broadening} we plot the strong-coupling results (red) and
compare them with the non-cavity ones (blue) with (light) and without (dark)
state broadening. Figures \ref{fig:broadening}a and b show the results for a
system with an $S_1-TT$ gap of $0.2$ eV, representative of the class of
materials we propose for cavity-mediated singlet fission. Like in the previous
sections, the coupling strength is chosen such that the $\mathbf{LP}$ is close
to resonance with the $\mathbf{TT}$ eigenstates ($\Omega_R = 375$ meV). 
In general, the singlet-fission dynamics in the cavity is slowed down
once state broadening is considered (light vs dark), but it is still much faster than
the non-cavity situation (red vs blue). Therefore, our proposal is still operative 
in a realistic situation where the states have a finite width, and strong
coupling can enhance singlet fission dynamics of materials that present a large
$S_1-TT$ gap even in the presence of state broadening. 

We explore next the situation in which the $S_1-TT$ gap is smaller ($0.08$ eV) and, accordingly, 
the Rabi splitting is also reduced ($0.15$ eV) to match the $\mathbf{LP}$ spectral location with that of the $\mathbf{TT}$ eigenstate. 
Our results are shown in Figures \ref{fig:broadening}c and d for the two 
sampling cases. Note first that when comparing the dark lines (the same in both panels), 
in which state broadening is not taken into account, the cavity and non-cavity dynamics (red vs blue)
are more alike than for the case with a larger $S_1-TT$ gap (Figures
\ref{fig:broadening}a and b). As discussed before, this is 
due to the rapid relaxation to the dark state manifold when the $S_1-TT$ gap is small.
Moreover, when state broadening is incorporated in the simulations, 
cavity dynamics (light red line) turns out to be very similar to the non-cavity
situation (light blue line) and the shaded areas even overlap.
These results nicely explain why in a recent experimental work \cite{menon2020} no significant
differences were found when comparing the cavity and non-cavity singlet fission dynamics of TIPS-pentacene. 
In the experimental setup, both the Rabi splitting and the FWHM of the $S_1$ vibronic peak that is coupled to the
cavity mode were $\sim 0.1$ eV.  Notably, the $S_1-TT$ gap in
TIPS-pentacene is also expected to be close to this value. \cite{PhysRevLett.115.107401}
Therefore, the experimental conditions resemble those illustrated in Figures
\ref{fig:broadening}c and d, which predict very moderate changes in singlet fission dynamics.
This further confirms that singlet fission rates in slightly exothermic systems are not expected 
to be largely boosted by strong coupling.

\section{Conclusions}
To conclude, we have demonstrated that singlet fission dynamics can be
accelerated under collective strong coupling via state-mixing when the lower
polariton is almost resonant with the $TT$ state. We have also shown that this effect is
much more beneficial for compounds with large $S_1-TT$ gaps, thus reducing both the
population transfer from the $\mathbf{LP}$ to the $\mathbf{S_1}$ dark state manifold and the detrimental 
impact that energetically broadening has on the resonant mechanism behind the rate 
acceleration. Given that the main characteristic of conventional singlet-fission materials is a small $S_1-TT$ gap,
and that systems such as polycyclic aromatic hydrocarbons are unstable and pose
significant challenges for practical applications, our results provide a new
perspective and suggest a new paradigm for cavity-mediated singlet fission with
materials that have not previously been considered for singlet fission. We hope
that our results can serve as a guide and inspire future experiments to realize 
efficient singlet fission with unconventional compounds in optical cavities.

\section{Acknowledgements}
We are grateful to Dr.~Rocío Sáez Blázquez for advice regarding the 
Bloch-Redfield simulations and also for interesting discussions.
This work has been funded by the European Research Council through Grant ERC-2016-StG-714870 
and by the Spanish Ministry for Science, Innovation, 
and Universities -- Agencia Estatal de Investigación through grants 
RTI2018-099737-B-I00, PCI2018-093145 (through the QuantERA program of the European Commission), 
and MDM-2014-0377 (through the María de Maeztu program for Units of Excellence in R\&D). 
DC acknowledges financial support from the Ministerio de Economía y Competitividad  
(Projects PID2019-109555GB-I00 and RED2018-102815-T) and the Eusko Jaurlaritza (Project PIBA19-0004). 

\section{Methods}

In our simulations each molecular excited state 
$\{ |TT_k \rangle, |S_{1_k} \rangle, |CT_k \rangle\}$  
is coupled to identical and independent baths. We 
consider Ohmic spectral densities with a Lorentizian cutoff 
$ J(w)=2\lambda \Omega w/(w^2 + \Omega^2)$, where $\Omega$ is the cutoff frequency 
and $\lambda$ the reorganization energy. Following reference \citenum{reichmanI}, 
we take the values $\Omega=150$ meV and $\lambda=25$ meV, which are characteristic of 
organic aromatic molecules.

The power spectrum is given by
\begin{equation}
    S(w)=
    \begin{cases}
        2J(w)(n(w,T)+1) &, w>0 \\
        4kT\lambda/\Omega &, w=0 \\
        2J(-w)n(-w,T) &, w<0
    \end{cases}
\end{equation}
with the Bose occupation factor $n(w,T)=(e^{w/k_B T} -1)^{-1}$, 
and we take $T=300$ K.

To account for the finitie cavity lifetime we include a Lindblad term in our simulations, 
$\frac{\kappa}{2} \mathcal{L}_{\hat{a}}[\hat \rho]$, where 
$\mathcal{L}_{\hat{a}} = 2 \hat{a}\hat{\rho}\hat{a}^\dagger - 
\{\hat{\rho},\hat{a}^\dagger\hat{a}\}$, where $\hat{a}$ is the bosonic destruction 
operator of the cavity photon and $\kappa$ is the lifetime.

Note that we do not rely on the secular approximation of 
the Bloch-Redfield equation. This is because this approximation fails when 
there are eigenstates close in energy. The master equation has been solved with the 
Qutip package. \cite{qutip2012,qutip2013}


\bibliography{main}

\newpage

\begin{figure}[H]
  \includegraphics[width=100mm]{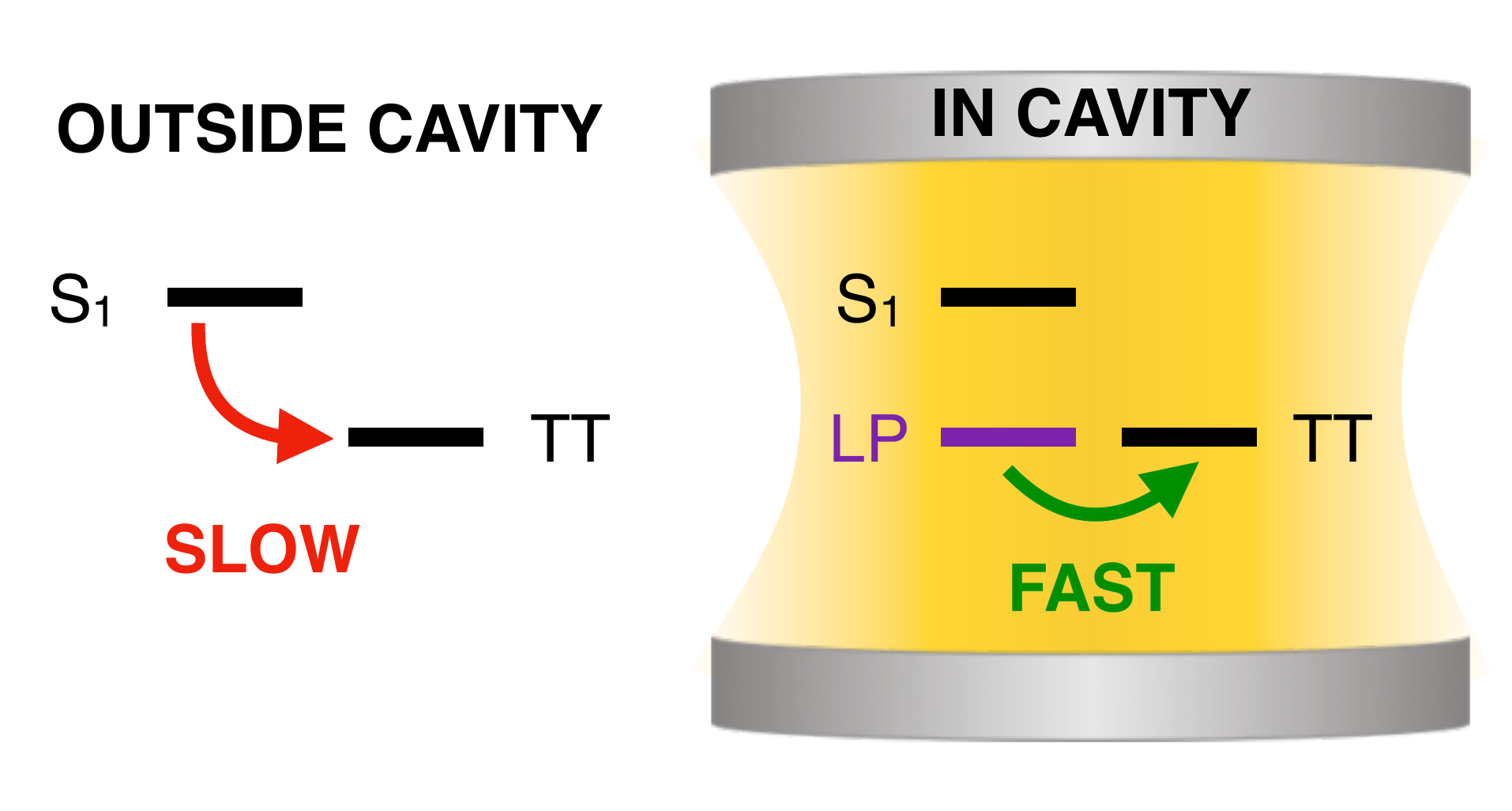}
  \caption*{For Table of Contents Only}
\end{figure}
\end{document}